\documentclass[aps,prl,twocolumn,amsmath,amssymb,nofootinbib]{revtex4-2}
\usepackage{graphicx}
\usepackage{hyperref}
\usepackage{braket}
\usepackage{mathtools}
\usepackage{amsfonts}
\usepackage{physics}
\usepackage{bm}
\usepackage{xcolor}

\begin{document}

\title{Extracting Edge Modes: Reduction of 3D and 2D Gravities}
\author{Euihun Joung}
\author{Sejin Kim}
\affiliation{Department of Physics, Kyung Hee University, Seoul 02447, Korea}
\date{\today}

\begin{abstract}

We investigate the boundary reduction of 3D Einstein gravity and JT gravity into their respective 
edge mode theories—namely, the Liouville and Alekseev-Shatashvili, and Schwarzian models.
By examining the roles of boundary conditions, canonical transformations, and differing formulations—metric versus Chern-Simons—we clarify how physical degrees of freedom become localized at the boundary and resolve several long-standing ambiguities in the reduction procedure.

\end{abstract}

\maketitle

In gauge theories on manifolds with boundaries, gauge symmetries are typically broken at boundaries, where would-be gauge redundancies acquire physical meaning and manifest as genuine physical degrees of freedom.
These boundary degrees of freedom—commonly known as would-be gauge modes or edge modes—are frequently overlooked, as their contributions to physical observables are generally considered negligible compared to those from the bulk.
Given the recent surge of interest in AdS/CFT and defect field theories, it is timely to re-examine the role of edge modes, whose contributions could be critical to understanding boundary-related phenomena.

In topological field theories—such as Chern-Simons (CS) or BF theory—where no local propagating degrees of freedom exist in the bulk of the manifold, the system’s dynamical content is entirely encoded in its edge modes.
Notable examples are the free chiral scalar or chiral WZW theory in two dimensions, which can be obtained from Abelian and non‑Abelian CS theories by Hamiltonian reduction \cite{Floreanini:1987as,Witten:1988hf,Elitzur:1989nr}: these edge modes have played a key role in understanding various condensed‑matter properties, such as the quantum Hall effect \cite{Wen:1990kqa,Wen:1992vi}.
Edge modes in topological gauge theory can be generalized to higher dimensions \cite{Moore:1989yh,Arvanitakis:2022bnr,Chen:2025xlo}.

In this Letter, we revisit edge modes arising in three-dimensional gravity and Jackiw–Teitelboim (JT) gravity.
These gravity theories are topological—there are no propagating bulk degrees of freedom—and they can be reformulated as 3D CS and 2D BF theories, respectively.
We regard these low-dimensional models as a stepping stone toward uncovering gravitational edge modes in higher dimensions, and accordingly, we develop a framework that is readily extensible.
Before entering the details, let us review the vast literature related to this subject.\footnote{Our list of references is not complete; this is partly due to our limited knowledge and partly due to space constraints.} 
The role of edge modes in gravity was first noted by Balachandran et\,al.\ \cite{Balachandran:1994up}, who found the algebra of boundary charges using the symplectic structure of the edge modes.
\footnote{See also \cite{Carlip:1994gy,Carlip:1996mi,Carlip:1996yb} for the attempts of explaining the entropy of BTZ black hole in terms of the would-be gauge mode.}
These results are closely related to the asymptotic symmetries of gravity \cite{Brown:1986nw}, which were among the precursors of AdS/CFT. 
In modern terms, these derivations can be obtained most straightforwardly using the CS formulation of gravity. In treating gravity as a CS theory, the standard Hamiltonian reduction method of CS gauge theory 
can be used. Notably, by combining the two copies of $SL(2,\mathbb{R})$ WZW actions—which arise from the Hamiltonian reduction of 3D CS gravity with a specific boundary term—into a single $SL(2,\mathbb{R})$ WZW action and imposing constraints that amount to the asymptotic AdS condition, Coussaert et\,al.\ showed that one obtains Liouville theory \cite{Coussaert:1995zp},
\begin{equation}
	S=\frac{\ell}{32\pi G}\int_{\partial\mathcal{M}}{\rm d}^{2}x \bigg(\frac{1}{2}\partial_{\mu}\varphi\partial^{\mu}\varphi-8\;e^{\varphi}\bigg)\,,
\end{equation}
where $\varphi$ parameterizes the diagonal element of $SL(2,\mathbb R)$.
For more details, see \cite{Henneaux:1999ib,Rooman:2000zi} and the recent works \cite{Nguyen:2021pdz,Benizri:2023kjn}. 

Shortly after the work of \cite{Coussaert:1995zp}, a new perspective based on the AdS/CFT correspondence began to dominate studies of gravity, where one of the key objects is the on‑shell gravity action.
Working in the Fefferman–Graham gauge, the on‑shell reduction of gravity was systematically analyzed, notably by \cite{Henningson:1998gx,Imbimbo:1999bj,deHaro:2000vlm},
and it was understood how the breaking of radial diffeomorphisms is connected to the Weyl anomaly of the dual CFT—the holographic Weyl anomaly.
For the computation of this anomaly, one introduces the radial diffeomorphism in the on‑shell gravity action and interprets it as a \emph{wiggling} boundary position. The resulting position variable, a boundary scalar $\varphi$, is essentially a Goldstone boson and thus corresponds to the edge mode.
In the case of ${\rm AdS}_3$, it was observed that the anomalous energy‑momentum tensor of $\varphi$ takes the same form as the energy‑momentum tensor \cite{Navarro-Salas:1998fgp,Skenderis:1999nb,Bautier:2000mz} of the action,
\begin{equation}
	S=-\frac{\ell}{32\pi G}\int_{\partial\mathcal{M}}{\rm d}^2x \sqrt{-g} \bigg(\frac{1}{2}\partial_{\mu}\varphi\partial^{\mu}\varphi-R\varphi\bigg)\,.
\end{equation}
Many other works on holography, such as \cite{Nakatsu:1999wt,Rooman:2000ei,Krasnov:2000zq,Krasnov:2001cu,Manvelyan:2001pv}, have made similar observations, but 
it was in \cite{Carlip:2005tz} where the physical meaning of the Liouville action as part of the gravitational degrees of freedom was clearly articulated.  In that work, it was also pointed out that the two derivations of the Liouville action—the Hamiltonian reduction (Coussaert et\,al.) and the on-shell reduction (holography)—display a significant discrepancy.
The former includes the potential term $e^\varphi$ while the latter does not.
In fact, there is another crucial difference between the two actions that has not been addressed in the literature—the overall sign of the action.

These two computations use very different procedures, making it nontrivial to identify the origin of this discrepancy. One notable distinction is that they employ different boundary terms and boundary conditions.
The Hamiltonian reduction uses the typical boundary conditions and boundary terms standard in CS gauge theory, but these are rather odd in gravity because the boundary term $\mathrm{Tr}(A\wedge \star A)$ 
involves a flat metric through Hodge star $\star$ in addition to the genuine metric hidden inside $A$---More precisely, the typical boundary term in Hamiltonian reduction is what one obtains after integrating the $A_0$ components.
It also employs a gauge choice that is unnatural from a gravitational point of view: rather than solving the algebraic torsion constraint, it solves differential constraints. The computation can be performed whether the boundary lies at finite distance or at infinity; the asymptotic nature of the boundary is enforced through additional constraints that ultimately transform the WZW action into Liouville theory.
In the holographic reduction, one uses the metric formulation and employs the Gibbos-Hawking-York terms with suitable counterterms as boundary terms and imposes Dirichlet boundary conditions, working in the Fefferman–Graham gauge.
Here, the on‑shell action diverges, and we need to perform holographic renormalization: we set the boundary at a finite distance, compute everything with counterterms, and then send the boundary to infinity at the end. Despite the remarks of \cite{Carlip:2005tz}, these discrepancies went unaddressed for about two decades.

In recent years, gravitational edge modes have attracted renewed interest due to major progress made by the duality between the SYK model and JT gravity \cite{Maldacena:2016hyu,KitaevTalks1,*KitaevTalks2,Kitaev:2017awl}. This two‑dimensional theory closely parallels the three‑dimensional case. JT gravity is topological and can be reformulated as an $SL(2,\mathbb{R})$ BF gauge theory. Maldacena et\,al.\ first showed that JT gravity with a wiggling boundary can be reduced to the Schwarzian theory \cite{Maldacena:2016upp}; the same conclusion was later drawn from a partial on-shell evaluation of the BF theory \cite{Saad:2019lba}. 
Although the analyses of JT gravity are closely analogous to those of 3D gravity, no discrepancy analogous to the potential‑term issue identified in \cite{Carlip:2005tz} has been noted.

More recently, this reduction scheme has been applied to 3D gravity by Cotler and Jensen \cite{Cotler:2018zff}.
Using Hamiltonian reduction but, instead of merging two WZW actions, separately imposing the asymptotic AdS condition on each of the chiral WZW actions, they obtained the Alekseev–Shatashvili‑type (AS) action \cite{Alekseev:1988ce} and argued that it is the correct edge‑mode action of 3D gravity rather than the Liouville theory.

In both the Schwarzian and AS theories, the temperature‑sensitive quadratic term is analogous to the Liouville potential; these terms are introduced through a reparameterization trick.
Later, it was understood that this term was originally absent because the Gauss decomposition used was compatible only with the zero‑temperature case. By using the Iwasawa decomposition, it was shown \cite{Valach:2019jzv} that the temperature‑dependent quadratic term can be obtained directly from the BF or CS analysis.
This observation raises another puzzle; 
the original derivation \cite{Coussaert:1995zp} of the Liouville theory with potential term  actually used the Gauss decomposition, seemingly at odds with the AS theory \cite{Cotler:2018zff}. 

To analyze all of these issues in a systematic manner, we adopt the following methods.
First, we work with the radial Hamiltonian method, in which the radial coordinate $r$ is treated as the Hamiltonian evolution parameter. This brings the gravity action into an ADM‑like form, but the role of time $t$ is played by $r$.
This method is commonly used in the on‑shell reduction of gravity theories in AdS/CFT (see e.g. \cite{deBoer:1999tgo,deHaro:2000vlm}) and is sometimes even generalized to a Hamilton–Jacobi formulation \cite{Papadimitriou:2004rz}. In this setup, issues of boundary conditions and boundary terms are controlled by canonical transformations, since the boundary terms are nothing but generating functions of canonical transformations. The role of canonical transformations in holographic renormalization, i.e., the counterterms, was explored in \cite{Papadimitriou:2010as}.
Our viewpoint is a bit more general: we regard the canonical transformation as a tool to control the most general boundary conditions and boundary terms of a theory—that is, the most general edge‑mode theory. The related issues require more detailed explanations and will be treated in a separate paper.

Here, we take the following Hamiltonian action as our starting point.
 \begin{equation}\label{eq:radial Hamiltonian CS}
	\begin{aligned}
		&S_{\rm ren}[A]
		=S_{\rm CS}[A]-\frac{\ell}{16\pi G}\int_{\partial \cal M}(\epsilon_{\alpha\beta}\,{e}^{\alpha}\wedge f^{\beta}-b\wedge \omega)\\
		&=-\frac{\ell}{8\pi G}\int_{\mathcal{M}}{\rm d}r \bigg[\epsilon_{\alpha\beta}f^{\alpha}\wedge\partial_{r} e^{\beta}-b\wedge\partial_{r}\omega\\
		&\qquad\qquad
		+\epsilon_{\alpha\beta}\left(
		n^{\alpha}\left({\rm D}f^{\beta}-b\wedge f^{\beta}\right)
		-m^{\alpha}\left({\rm D} e^{\beta}+b\wedge  e^{\beta}\right)\right)\\
		&\qquad\qquad -n\left({\rm d}\omega+\epsilon_{\alpha\beta}f^{\alpha}\wedge e^{\beta}\right)
		-m\left({\rm d} b+e^{\alpha}\wedge f_{\alpha}\right)\bigg]\,,
	\end{aligned}
\end{equation}
 where ${\rm D}V^\alpha={\rm d}V^\alpha+\omega\wedge \epsilon^{\alpha\beta}\,V_\beta$ is the 2D Lorentz  covariant derivative and $\alpha, \beta=0,1$.
Here, the level of Chern-Simons action is chosen as $k=\frac{\ell}{4\pi G}$.
The $so(2,2)$ connection $A$ is decomposed as
 \begin{equation}
	\begin{multlined}
	A=( e^{\alpha}+n^{\alpha}{\rm d}r)P_{\alpha}+(f^{\alpha}+m^{\alpha}{\rm d} r)K_{\alpha}\\
	+(\omega+m\,{\rm d}r)\,J+(b+n\,{\rm d}r)\,D\,,
	\end{multlined}
\end{equation}
in terms of the generators of conformal algebras with non-trivial Lie brackets,
\begin{equation}
	\begin{aligned}
	&[P_\alpha, D]=+P_\alpha\,, \qquad [P_\alpha, J]=\epsilon_{\alpha\beta}\,P^\beta\,,\\
	&[K_\alpha, D]=-K_\alpha\,, \qquad [K_\alpha, J]=\epsilon_{\alpha\beta}\,K^\beta\,,\\
	& [P_\alpha, K_\beta]=\eta_{\alpha\beta}D+\epsilon_{\alpha\beta}J\,.
	 \end{aligned}
\end{equation}
Here, $e^\alpha$, $f^\alpha$, $\omega$, and $b$ are one‑forms on constant‑$r$ hyperplanes,
while the Lagrange multipliers $n^\alpha$, $n$, $m^\alpha$ and  $m$ are scalars.
Note here that the system of constraints is nothing but the frame formulation of conformal geometry,
see e.g. \cite{Joung:2021bhf} for a review:  
In the $b=0$ gauge, we have the usual torsionless constraint, ${\rm D}e^\alpha=0$.
The constraints $e^\alpha\wedge f_\alpha=0$ and ${\rm d}\omega+\epsilon_{\alpha\beta}f^\alpha\wedge e^\beta=0$ tell that $f_{\mu\nu}=e_{\alpha\mu}\,f^\alpha{}_\nu$ is the Schouten tensor,
and the constraint ${\rm D}f^\alpha=0$ tells that the Cotton tensor vanishes.
In the current case of 2D conformal geometry, the Cotton tensor vanishes identically, hence
the corresponding constraint is a consequence of the others.

Note that our boundary term differs from $\mathrm{Tr}(A\wedge \star A)$ used in the usual CS formulation:
this term preserves $so(2,2)$ gauge symmetry but breaks diffeomorphisms,
whereas our boundary term preserves diffeomorphisms but breaks a part of $so(2,2)$.
However, for the flat boundary metric $e^\alpha=e^{r}{\rm d}x^{\alpha}$,
the two boundary terms coincide. Therefore our boundary term can be viewed as 
a geometric generalization of the usual boundary term of CS formulation.

As the subscript of $S_{\rm ren}$ hints, our action already contains all necessary counterterms 
via a suitable canonical transformation that brings the gravity action
into the above form.
The variation of the action gives 
\begin{equation}
	\delta S_{\rm ren}=({\rm EoM})-\frac{\ell}{8\pi G}\int_{\partial\cal M} \epsilon_{\alpha\beta} f^\alpha
	 \wedge\delta  e^\beta
	-b\wedge\delta \omega\,,
\end{equation} 
 and we choose the boundary conditions $\delta  e^\alpha=0$
 and $b=0$. Under the $so(2,2)$ gauge symmetry,
\begin{equation}
	\delta S_{\rm ren}=\frac{\ell}{8\pi G}\int_{\partial\cal M} 
	(\epsilon_{\alpha\beta} \kappa^\alpha {\rm D} e^\beta
	+\sigma\,{\rm d}\omega)\,,
\end{equation}
where $\kappa^\alpha$ and $\sigma$ are respectively the $K_\alpha$ and $D$ gauge 
transformation parameters, and hence those parts of gauge symmetry are broken at the boundary.
Under a finite $\kappa^\alpha$ transformation, $b$ transforms as $b\to b+ e^\alpha\,\kappa_\alpha$.
Since we set $b=0$ at the boundary, even though $\kappa_\alpha$ is broken at the boundary, we still have the (optional) freedom to gauge‑fix $b=0$ in the bulk using the unbroken bulk part of the $\kappa_\alpha$ gauge symmetry; this amounts to imposing the Fefferman–Graham gauge.
In the end, the $D$ gauge symmetry is broken at the boundary, and the corresponding gauge parameter $\sigma$ becomes the edge mode.

To isolate the edge‑mode action, we use the Stueckelberg transformation 
$A=g^{-1}(\mathrm{d}+\tilde A)\,g$, which renders a flat boundary into a wiggling one. 
With $g=e^{\sigma D}\,e^{\kappa^\alpha K_\alpha}$ at the boundary, we require the boundary condition $b=0$ to be preserved even after the Stueckelberg shift: $\tilde b=0$. 
This fixes $\kappa^\alpha$ in terms of $\sigma$ as $\kappa_{\alpha}[\sigma] = -e^{-\sigma}\,\tilde e_{\alpha}{}^{\mu}\,\partial_{\mu}\sigma$ and yields
\begin{equation}\label{Liouville rel}
	\begin{aligned}
		 e^{\alpha}&=e^{\sigma}\tilde{e}^{\alpha}\,,\\
		f^{\alpha}&=e^{-\sigma}\tilde{f}^{\alpha}+\tilde{\rm D}\kappa^{\alpha}-\frac{1}{2}e^\sigma\,\kappa^{\beta}\kappa_{\beta}\,\tilde e^{\alpha}\,,\\
		\omega&=\tilde{\omega}+e^\sigma\,\epsilon_{\alpha\beta}\,\tilde e^{\alpha}\kappa^{\beta}\,.
	\end{aligned}
\end{equation}
Using the finite gauge transformation of CS action,
\begin{equation}
	\begin{aligned}
	S_{\rm CS}[A]&=S_{\rm CS}[\tilde A]-\frac{\ell}{16\pi G}\int_{\partial \cal M} \tr(g^{-1}{\rm d}g \wedge \tilde A )\\
	&\qquad\qquad\; -\frac\ell{48\pi G}\int_{\cal M}\tr(g^{-1}{\rm d}g)^3, 
	\end{aligned}
\end{equation}
we find 
\begin{equation}\label{eq:Liouville offshell}
	S_{\rm ren}[A]=S_{\rm ren}[\tilde{A}]+\frac{\ell}{16\pi G}\int_{\partial\mathcal{M}}
	\sqrt{-\tilde g}\left(-\tilde g^{\mu\nu}\,\partial_{\mu}\sigma\partial_{\nu}\sigma-\sigma \tilde R\right).
\end{equation}
Above, we solved the torsion constraint $\tilde{\mathrm{D}}\tilde e^\alpha=0$ to write the final edge‑mode action in metric form. 
By construction, the above action is invariant, aside from all other gauge transformations, under the $D$ Stueckelberg transformation,
\begin{equation}
	\sigma\to \sigma+\varsigma\,, \qquad \tilde A\to {g_\varsigma}^{-1}({\rm d}+\tilde A)g_\varsigma\,,
\end{equation}
with $g_\varsigma=e^{\varsigma\,D}\,e^{K_\alpha\,\kappa^\alpha[\varsigma]}$\,.

If we solve the bulk equations for $\tilde A$, the bulk part $S_{\rm ren}[\tilde A]$
does not contain any $\sigma$ dependence and reduces to zero, 
up to the linear divergence term in $r$ proportional to $\int_{\partial \mathcal M} 
\epsilon_{\alpha\beta}\tilde f^{\alpha}\wedge \tilde e^{\beta}$,
which becomes a boundary term $\int_{\partial \mathcal M} {\rm d}\omega$ by constraint.
The remaining action is the Liouville action but without any exponential potential term.
This exactly reproduces the result obtained in the holographic method \cite{Skenderis:1999nb}
and leads to
the puzzle of discrepancy between the two methods as remarked in \cite{Carlip:2005tz}
and the begining of this Letter.

In this paper, we adopt a new boundary term in the CS formulation to bridge the previously studied two methods. We use the Stueckelberg formulation, which is equivalent to introducing a wiggling boundary; however, 
the distinction between the original field $A$ and its Stueckelberg‑shifted counterpart $\tilde A$ is kept explicit.
This allows us to make the following observations.

The first observation is that the boundary condition should be imposed on the original connection $A$ rather than on $\tilde{A}$. Accordingly we must express $\tilde{e}^\alpha$ in terms of $e^\alpha$ and $\sigma$
at the boundary. This brings the action into the form
\begin{equation}\label{eq:Liouville offshell'}
    S_{\rm ren}[A]=S_{\rm ren}[\tilde{A}]+\frac{\ell}{16\pi G}\int_{\partial\mathcal{M}}
    \sqrt{-g}\left(g^{\mu\nu}\,\partial_{\mu}\sigma\partial_{\nu}\sigma-\sigma R\right)\,,
\end{equation}
whose only difference from \eqref{eq:Liouville offshell} is that the kinetic term for the edge mode now has a positive sign. This sign is the correct one from the unitarity perspective as well as in comparison with the result obtained from the Hamiltonian reduction. Thus, imposing the boundary condition on the correct field fixes the sign but does not generate a potential term.

The second observation is that the on-shell reduction of $S_{\rm ren}[\tilde{A}]$ becomes more delicate
due to the boundary condition imposed on $A$.
If we solve all the bulk equation of motion $\tilde F=0$ and extend it to boundary, the boundary condition 
subjects also the edge mode $\sigma$  on-shell. In order to keep the edge mode off-shell ---
which is important for the quantization of the theory--- we should not impose 
$\tilde F_{0r}=0$ and the $D$ part of $\tilde F_{01}=0$, at least close enough to the boundary.
Even with a part of equations, $S_{\rm ren}[\tilde A]$ is reduced  to a boundary action.
Gauge fixing the Lagrange multipliers $\tilde n^\alpha, \tilde m^\alpha, \tilde m$ to zero and $\tilde n=1$---which amounts to impose the asymptotic AdS condition---
the equations $\tilde F_{1r}=0$ become as simple as
\begin{equation} \label{bulk eq}
	 \partial_r\tilde e^\alpha_1=\tilde e^\alpha_1\,, 
	\quad \partial_r\tilde f^\alpha_1=-\tilde f^\alpha_1\,,\quad  \partial_r \tilde b_1=0\,,\quad 	\partial_r\tilde\omega_1=0\,.
\end{equation}
Hence, only the radial dependence of the ${\rm d}x^1$ component is fixed.
If we had imposed all the equations, the radial dependence of the ${\rm d}x^0$ components would also be fixed. In that case the $D$ component of $\tilde F_{01}=0$,
${\rm d}\tilde \omega+\epsilon_{\alpha\beta}\,\tilde f^\alpha\wedge \tilde e^\beta=0$,
becomes the equation of motion for the edge mode.
With this gauge choice and the equations \eqref{bulk eq}, the bulk part of the action is reduced to
$S_{\rm ren}[\tilde A]
	=\tfrac{\ell}{8 \pi G}\int_{\partial \cal M} {\rm d}^2x\,
	\epsilon_{\alpha\beta}
	\tilde f_1^\alpha\,\tilde e^\beta_0$,
up to the linearly divergent term in $r$ proportional to the boundary term $\int_{\partial\cal M} {\rm d}\tilde\omega$.
For simplicity, we impose a flat two‑dimensional metric boundary condition, $e^\alpha=e^r\,{\rm d}x^\alpha$,
then  the action becomes
$S_{\rm ren}[\tilde A]=-\tfrac{\ell}{8\pi G}\int_{\partial \mathcal{M}} {\rm d}^2x\,\mu(x)\,e^{-\sigma}$.
Here, $\mu(x)$ given by $\tilde f^1{}_1(r,x)=e^{-r}\,\mu(x)$ can be set to a constant
by the Stueckelberg symmetry, $\mu\ \to \ 
	e^{-\varsigma}\,(\mu
	-(\partial_1)^2\varsigma+
	\partial_1\varsigma\,\partial_1\varsigma-\frac12\,
	\partial^\rho\varsigma\,\partial_\rho\varsigma)$.
Constant $\mu$ cannot be removed as it would require a physics-changing \emph{large} Stueckelberg transformation.
Eventually, we end up with the Liouville action,
\begin{equation}
	S_{\rm ren}[A]\simeq \frac{\ell}{16\pi G}
	\int_{\partial\cal M} {\rm d}^2x\left[\partial_{\mu} \sigma\partial^{\mu}\sigma-2\,\mu\,e^{-\sigma}\right],
\end{equation}
where $\simeq$ stands for the fact that we solved a part of bulk equations of motion.
Remark that the above procedure breaks the manifest 2D Lorentz symmetry in the intermediate step,
but it is also possible to devise an improved procedure which preserve Lorentz symmetry manifestly \cite{Arvanitakis:2022bnr, Evnin:2023ypu}.

In the above derivation, we introduced Stueckelberg fields 
only for the broken $K_\alpha$ and $D$ gauge symmetries, that is, the three fields $\kappa^\alpha$ and $\sigma$.
Requiring the two conditions of $\tilde b_\mu=0$, we end up with only one edge mode
$\sigma$.
If we had included $P_\alpha$ gauge symmetries with the 
corresponding Stueckelberg fields $\xi^\alpha$, it simply leads to the 
boundary action transformed by diffeomorphism $\xi^\mu$,  which could be cancelled 
by transforming also the boundary condition, that is, the boundary metric.

Next, we consider a different reduction procedure. 
The decomposition $so(2,2)=sl(2,\mathbb R)_{\rm L}\oplus sl(2,\mathbb R)_{\rm R}$,
where each $sl(2,\mathbb R)_{\rm L/R}$ is generated by $P_\pm, K_\mp, D_\pm=D\pm J$
can lead to the decomposition of the action itself, 
$S_{\rm ren}[A=A^{\rm L}+A^{\rm R}]=S_{\rm ren}[A^{\rm L}]-S_{\rm ren}[A^{\rm R}]$,
thanks to the boundary condition $b=0$.
Focussing the half of the action $S_{\rm ren}[A^{\rm L/R}]$, we introduce Stueckelberg fields for all three gauge symmetries while imposing two conditions $\tilde{b}_1\pm \tilde{\omega}_1=0$ and $\tilde{e}^\pm_1 = \pm\,e^r$. This leaves one edge mode at each half of the action and yields the chiral action of Alekseev–Shatashvili type \cite{Cotler:2018zff,Alekseev:1988ce}. 

We introduce these Stueckelberg fields via the finite transformation,
\begin{equation}\label{pure gauge}
    g_\pm=e^{\xi^\pm\left(e^r P_\pm+e^{-r}\mathcal{L}_\pm K_\mp\right)}
    e^{\sigma_\pm D_\pm}\,e^{e^{-r}\kappa^\mp K_\mp}\,,
\end{equation}
which corresponds to the background metric
${\rm d}s^2={\rm d}r^2-(e^{r}{\rm d}x^{+}+e^{-r}\mathcal{L}_{-}{\rm d}x^{-})(e^{r}{\rm d}x^{-}+e^{-r}\mathcal{L}_{+}{\rm d}x^{+})$ where $x^{\pm}=\tfrac{1}{2}(x^{0}\pm x^{1})$.
The conditions $\tilde{b}_1\pm\tilde{\omega}_1=0$ and $\tilde{e}^\pm_1 = \pm e^r$ yield
\begin{equation}
\begin{aligned}\label{AS rel}
	&e^{-\sigma_\pm}=
	1\pm \partial_1\xi^\pm\\
	&e^{r}\,f^\pm{}_1=
	e^{-\sigma_\pm}\,{\cal L}_\pm(\mp1+ \partial_{1}\xi^\pm)+\partial_1\kappa^\pm\mp(\kappa^\pm)^2\,,
	\\
	&0=\partial_1\sigma^\pm\mp 2\kappa^\mp\,,
\end{aligned}
\end{equation}
along with analogous ${\rm d}x^0$ components that involve unfixed quantities such as $\tilde{e}^\pm{}_0$ and $\tilde{f}^\pm{}_0$. Note also that we have chosen $\tilde f^\pm{}_1=\mp e^{-r}\,\mathcal{L}_\pm$, which can be achieved by the $D_\pm$ Stueckelberg transformation. The above relations determine $\kappa^\pm$ in terms of $\partial_1\sigma^\pm$ and $\sigma^\pm$ in terms of $\partial_1\xi^\pm$. Substituting these into the action $S_{\rm ren}[A^{\rm L/R}]$ yields
\begin{equation}
\begin{aligned}
	& S_{\rm ren}[A^{\rm L/R}]=S_{\rm ren}[\tilde A^{\rm L/R}]
	-\frac{\ell}{8\pi G} \int_{\partial \cal M}{\rm d}^2x\,\tilde f^\mp_1\,\tilde e^\pm{}_0
	\\
	& 
	\mp\frac{\ell}{16\pi G}\int_{\partial \cal M}{\rm d}^2x\bigg[\frac{
	\partial_{\mp}\partial_{1}\chi^{\rm L/R} \partial_{1}^2\chi^{\rm L/R}}{(\partial_{1}\chi^{\rm L/R})^2}+4\mathcal{L_\mp}
	\partial_{\mp}\chi^{\rm L/R}\partial_{1}\chi^{\rm L/R}\bigg],
\end{aligned}	
\end{equation}
where the second line is the AS‑like action $\pm S^{\rm L/R}_{\rm AS}[\chi^{\rm L/R}]$ found in \cite{Cotler:2018zff}, and $\chi^{\rm L/R}=\pm\varphi+\xi^\pm$ is the edge mode.

A few remarks are in order. First, this action already contains a potential term, namely the $\mathcal{L}_\mp$‑dependent quadratic term, even though we have not yet reduced the bulk part. Second, in addition to the bulk part, the first line includes an extra boundary term $\tilde f^\mp_1\,\tilde e^\pm{}_0$, so that the first line can be written as $S_{\rm CS}[\tilde A^{\rm L/R}]+\tfrac{\ell}{8\pi G}\int_{\partial \mathcal{M}} \tilde f^\mp{}_0\,\tilde e^\pm{}_1$. Third, the on‑shell reduction of this first line vanishes upon substituting the pure‑gauge solution \eqref{pure gauge}; using a more general solution produces an additional edge mode that eventually combines with $\chi^{\rm L/R}$. Finally, we confirm that the 3D gravity reduces to two copies of AS theories:
\begin{equation}
	S_{\rm ren}[A]\simeq S^{\rm L}_{\rm AS}[\chi^{\rm L}]-S^{\rm R}_{\rm AS}[\chi^{\rm R}]\,.
\end{equation}

We have seen that the 3D gravity can be reduced both to Liouville and AS theory
by Stueckelberg shift and solving a part of bulk equations.
Since we can always work with the unitary gauge where Stueckelberg fields vanish,
the reduction to Liouville and AS theory is possible if one solves just the right part of bulk equations,
which  in fact contains  the constraint in the usual temporal Hamiltonian. Therefore
our derivation should be compatible with the Hamiltonian reduction: this relation is 
straightforward in the derivation of AS theory, but rather involved in the derivation of Liouville as 
one needs to go through steps which breaks manifest 2D Lorentz.
In our derivation, the potential term appears only in the case where the background bulk solution
admits non-trivial boundary topology. This is in contrast with the result \cite{Coussaert:1995zp}.

The JT gravity, which is the 2D analogue of the 3D gravity, is the playground where
all the above derivations can be repeated in an analogous but simpler way.
The starting point is the action,
\begin{equation}
	\begin{aligned}
		&S_{\rm ren}[A,\Phi]=S_{\rm BF}[A,\Phi]-\frac{\ell}{8\pi G}\int_{\partial \cal M}(\phi_{P} f
		-\phi_D\,b)\\
		&=\frac{\ell}{8\pi G}\int_{\mathcal M}{\rm d} r\bigg[\phi_{K}\partial_{r}  e-f\partial_{r}\phi_{P}
		-\frac{1}{2}b\,\partial_r\phi_{D} \\
		&\qquad +n({\rm d}\phi_{K}+\phi_{K}b-\phi_{D}f)\\
		&\qquad +m({\rm d}\phi_{P}-\phi_{P}b+\phi_{D} e)\\
		&\qquad +\frac{1}{2}l({\rm d}\phi_{D}+2\phi_{P}f-2\phi_{K} e) \bigg]\,,
	\end{aligned}
\end{equation}
where $S_{\rm BF}[A,\Phi]=\tfrac{\ell}{16\pi G}\int_{\cal M}\tr (\Phi\;F)$ with $F={\rm d}A+A\wedge A$, and the $sl(2,\mathbb R)$ gauge connection $A$ and scalar field $\Phi$ are given by
\begin{equation}
	\begin{aligned}
		&A=(e+n\,{\rm d} r)P+(f+m\,{\rm d} r)K+(b+l\,{\rm d} r)D\,,\\
		&\Phi=\phi_{P}P+\phi_{K}K+\phi_{D}D\,.
	\end{aligned}
\end{equation}
The variational principle works with
$\int_{\partial\cal M} \phi_K\,\delta e -f\,\delta\phi_P-\frac12\,b\,\delta\phi_D=0$.
We impose the boundary condition $\delta e=0=\delta \phi_P$ and $b=0$.
This action reproduces the JT gravity with a proper boundary term. 
The gauge variation gives
$\delta S_{\rm ren}=\int_{\partial\cal M} \kappa({\rm d}\phi_P+\phi_D\,e)
+\frac12\sigma\,{\rm d}\phi_D$ and we find $K$ and $D$ symmetries are broken
at the boundary.

Similar to the derivation of the Liouville action, we introduce Stueckelberg fields for the broken symmetry via $g=e^{\sigma D}\,e^{\kappa K}$ and impose $\tilde{b}=0$ at the boundary. This fixes $\tau=\tfrac12\,e^{-1}\,\partial_t\sigma$ and leads to relations analogous to the first two of \eqref{Liouville rel}. With this we find
\begin{equation}
    S_{\rm ren}=S_{\rm BF}[\tilde{\Phi},\tilde{A}]-\frac{\ell}{8\pi G}\int {\rm d}t\,\frac{\phi_{P}}{2\,e_{t}}\Bigl(\tfrac{1}{2}(\partial_t\sigma)^2+\partial_{t}^2\sigma+\tilde{f}_{t}\,e^{-\sigma}\Bigr)\,.
\end{equation}
Thus, even before reducing the action we obtain a Liouville‑like edge‑mode action with a potential term. Solving the full bulk equations of motion would put the edge mode on shell. We therefore impose only the equation $F=0$, which suffices to reduce the bulk term to zero.

A procedure analogous to the derivation of the AS‑like action is also available. We introduce all three Stueckelberg fields $\xi,\sigma,\kappa$ and impose the two conditions $\tilde{e}_{t}=e_{t}$ and $\tilde{b}=0$. Setting $e_t=e^r$ and $\tilde{f}=e^{-r}\mathcal{L}$ yields relations analogous to \eqref{AS rel} and produces a Schwarzian action,
\begin{equation}
    S_{\rm ren}[A,\Phi]\simeq -\frac{\ell}{8\pi G}\int {\rm d} t\,\varphi_{P}\left(\frac{\dddot{\chi}}{\dot{\chi}}-\frac{3}{2}\Bigl(\frac{\ddot{\chi}}{\dot{\chi}}\Bigr)^2-2\,\mathcal{L}\,\dot{\chi}^2\right),
\end{equation}
where $\chi=t+\xi$ and boundary condition $\phi_{P}=2\;e^{r}\varphi_{P}$.
This Schwarzian action differs from \cite{Maldacena:2016hyu, Saad:2019lba} by an overall sign, a convention that is not physical because it can be absorbed into a redefinition of the dilaton boundary value $\phi_{P}$. Nevertheless, the overall sign issue is analogous to three-dimensional Chern–Simons gravity, where the corresponding sign choice is physically meaningful.
Specifically, the wiggly-boundary prescription of \cite{Maldacena:2016hyu} is equivalent to imposing boundary conditions on Stückelberg-shifted variables rather than on the original fields,
and the BF formulation of \cite{Saad:2019lba} imposes the partial on-shell condition $F=0$ instead of solving the Hamiltonian constraints.
Because these procedures differ from the derivation used here, a different overall sign is obtained.
See also \cite{Joung:2023doq} for other discussion of subtleties in deriving the Schwarzian action from a wiggling boundary, where the boundary conditions are still imposed on Stückelberg-shifted variables.

We expect the analysis presented in this Letter to make manifest and to clarify subtleties in the derivation of gravitational edge-mode actions in three and two dimensions, and to provide insight into edge-mode descriptions of higher-dimensional gravity and higher-spin theories.

\paragraph{Acknowledgements.
We thank Thomas Basile and Junggi Yoon for collaborations in early stage and useful discussions. 
Our work is supported by the National Research Foundation of Korea (NRF) grant funded by the Korean government (MSIT) (No. 2022R1F1A1074977).}

\bibliographystyle{apsrev4-2}
\bibliography{list_of_ref}

\end{document}